\documentclass[12pt,a4]{article}
\def\beq{\begin{equation}}
\def\eeq{\end{equation}}
\usepackage[dvips]{graphicx}
\begin{document}
\title{A Graph Model for the quantum mechanics of a moving cyclic disturbance interacting at a spatial position}
\author{Daniel Brown}
\newcommand{\ictitle}
{A Graph Model for the quantum mechanics of a moving cyclic
disturbance interacting at a spatial position}

\newcommand{\icauthor}
{Daniel Brown\footnote{email: d.brown@cs.ucl.ac.uk}\\Depts of
Computer Science and Physics}
\newcommand{\icaddress}
{University College London, U.K.}

\begin{titlepage}
\begin{center}
{\large{\bf\ictitle}}
\bigskip \\ \icauthor \\ \mbox{} {\it \icaddress} \\ \today \\ \vspace{.5in}
\end{center}
\setcounter{page}{0}
\begin{abstract}

A statistical approach based on a directed cyclic graph, is used
to calculate the alternative positions in space and state of a
moving disturbance for a given observed time. The probability for
a freely moving entity interacting in a particular spatial
position is calculated and a formulation derived for the minimum
locus of uncertainty in position and momentum. This accords with
calculations for quantum mechanics. The model has proven amenable
to computer modelling.

\end{abstract}
\end{titlepage}
\section{Introduction}

An earlier paper\footnote{see Brown (2003)} detailed interrelated
connections between Space, State, alpha-time and beta-Time using a
directed cyclic graph. This lead to a potential locus in Space for
a defined Time Magnitude (comprising both alpha-time(rst') and
beta-time$(t^*)$) where:

\beq |T| = \sqrt{(nt^*)^2+(npst'+rst')^2} \eeq

The occurrence - at the State$\rightarrow$Space trigger (ph) - of
the bifurcation of identity to \emph{both} a change in
beta-time$(t^*)$ at an adjacent Space position, and the change in
alpha-time associated with its change in State was shown to result
in a fundamental ambiguity for a given magnitude of time: where an
entity is located in Space(nd) and what its State(rh) is. Since n
and r are variables, a range of alternative combinations of State
and Space positions can combine to form the same total time
magnitude $|T|$ from variable components of alpha-time and
beta-time. This can be represented for a fixed $|T|$ of magnitude
$|rst'|$ - assuming a null Space trigger point (ie. a photon) - as
a ``temporal arc" (see diagram 1
below):\\

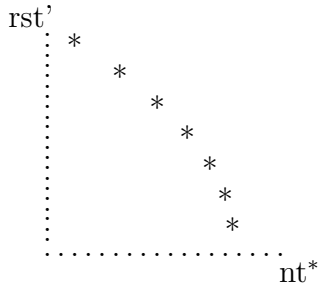
DIAGRAM 1 - temporal arc for a photon at a time magnitude $|rst'|$ \\
\hspace* {-6mm} rst'
\\[-1.6mm]
 \vdots\hspace* {2mm}*
\\[-1.6mm]
 \vdots\hspace* {8mm}*
\\[-1.6mm]
\vdots\hspace* {13mm}*
\\[-1.6mm]
\vdots\hspace* {17mm}*
\\[-1.6mm]
\vdots\hspace* {20mm}*
\\[-1.6mm]
\vdots\hspace* {22mm}*
\\[-1.6mm]
\vdots\hspace* {23mm}*
\\[-3.2mm]
\ldots\ldots\ldots\ldots\ldots\ldots \hspace* {23mm}
\\[-1.6mm]
 \hspace* {30mm}
nt$^{*}$

\textbf{All points on the temporal arc have same time magnitude
$|T|$.}

It was noted that we can represent time:
($\alpha$Time,$\beta$Time) as a complex vector. We use a notation
of beta-time($t^*$) as real and alpha-Time($\imath rst'$) as
imaginary: \beq \b{T} = n t^* + \imath (n p + r) st' \eeq or where
$z =( p + r/n)s$ : \beq \b{T} = n (t^* + \imath z t') \eeq

\section{The probability for a freely moving entity interacting in a particular spatial position}

For a defined time magnitude $|T|$ a range of possible
combinations of State and Space positions exist along its arc.
When the time magnitude measured is very small then specific state
and spatial positions cannot accurately be divined. A very small
$|T|$ is quite likely to be formed entirely from alpha-time
changes or entirely from beta-time changes.

\textbf{Since alternative possible compositions of
beta-time$(nt^*)$ and alpha-time($\imath rst'$) exist for a given
time magnitude $|T|$ then \emph{only} a probabilistic method can
reference the position in space and state of the IFE disturbance}.

Calculation of P(x) the probability of the IFE disturbance being
located (through an interaction) at a specific spatial position is
somewhat more intricate than might at first be expected.\\

There are two core issues to consider in the calculation:

1. the probability of an interaction occurring at a specific
spatial position and specific alpha-time.

2. The probability of \textit{non interaction} up to the point at
which interaction occurs.

We can therefore represent the total probability for an IFE being
found to be located in a specific position for a specific time as:
$P_T = P_L P_N$ where $P_L$ is the probability of the interaction
occurring in a particular alpha-time and beta-time and $P_N$ is
the probability that there has not been an interaction up to that
point.

To cover the first issue, assume that the probability of an
interaction in a specific given spatial position and alpha-time is
given by $A dx$ (where A may be dependent on x and t).

A toy model where $t^*=t'=dx=1$ illustrates the probability of the
IFE being found to interact with another IFE at a given spatial
position (e.g. x=1) with the time magnitude undefined.

At time $|t|=0$, $P_L(x=1)=0$\\
At time t=1 $P_T(x=1) = \frac{1}{2}A P_N$ where $P_N(x,t)$ is the
probability
 that an interaction has NOT occurred up to time $|t| = \sqrt{(nt^*)^2 + (rst')^2}$ (for x=1, t=1, $P_N=1$). This assumes that there is an equal likelihood of the IFE taking a path in alpha-time or beta-time.\\
At time $|t|=2$ $P_T(x=1) = \frac{1}{4}A P_N$\\

Different probabilities exist for the IFE being located at spatial
position x=1 for all possible time magnitude values. If the time
magnitude is undefined, then all these probabilities have to be
summed to establish the total probability of $x=1$..

To calculate $P_L (x,t) $, we consider the end result of the
movement to have been a random walk across a 2 dimensional time
plane. The probability for a spatial position x with undefined
time magnitude is the sum of a set of probabilities for all
possible alpha times in that spatial position for all possible
time magnitudes. Thus the probability that the IFE is located in
spatial position 1 for a time magnitude of 1 is equivalent to the
probability that the alpha time is 0 for a time magnitude of 1.
The probability that the IFE is located in spatial position 1 for
a time magnitude of 2 is equivalent to the probability that the
alpha time is 1 for a time magnitude of 2 etc...

The alpha-time can be considered to reach a value rst' through a
sequence of increments each $r_i$. Thus $rst'=\sum_{i=1}^R r_i
st'$

The probability of arriving at an alpha-time rst' in a particular
sequence is $(P_1 dr_1)(P_2 dr_2)(P_3 dr_3)...(P_R dr_R)$ where
$P_i$ is the probability density for a particular increment $r_i$
in alpha time.

The probability of arriving at rst' in any sequence requires that
all possible sequences are considered.

$P_L(rst') = \int\int^\infty_{-\infty}...\int (P_1 dr_1)(P_2
dr_2)(P_3 dr_3)...(P_R dr_R) $ with the restriction that $r<
\sum_{i=1}^R r_i <(r+dr)$

This restriction can neatly be encapsulated using the Dirac delta:

$P(rst') = \int\int^\infty_{-\infty}...\int (P_1 dr_1)(P_2
dr_2)(P_3 dr_3)...(P_R dr_R)  \delta(\sum_{i=1}^R r_i - r) $

i.e. $P(rst') = \int\int^\infty_{-\infty}...\int (P_1 dr_1)(P_2
dr_2)(P_3 dr_3)...(P_R dr_R) \frac{1}{2\pi}\int^\infty_\infty
e^{\imath q(\sum_{i=1}^R r_i -r)}  dq$

$=\frac{1}{2\pi} \int^\infty_{-\infty} e^{-\imath qr} dq
\int^\infty_\infty   (P_1 e^{\imath qr_1}  dr_1)
\int^\infty_\infty (P_2 e^{\imath qr_2} dr_2)
 \int^\infty_\infty   (P_3 e^{\imath qr_3}   dr_3)...\int^\infty_\infty  (P_R e^{\imath qr_R}  dr_R)$

ie. $P_L(rst')dr=\frac{1}{2\pi}\int^\infty_{-\infty} e^{-\imath
qr}dq \prod_{i=1}^R \int^\infty_{-\infty} (P_i dr_i)e^{-\imath
qr_i}dr_i$

$= \frac{1}{2\pi}\int^\infty_{-\infty} e^{-\imath qr}dq [
\int^\infty_{-\infty} (P_c dr_c)e^{-\imath qr_c}dr_c]^R$ if all
increments assumed the same.

Expanding $e^{\imath qr_c}$ in Taylor's series gives:

$\int^\infty_\infty (P_c dr_c)e^{-\imath qr_c}dr_c =
\int^\infty_{-\infty}  P_c (1+ \imath qr_c - \frac{1}{2}q^2 r_c^2+
...)dr_c = 1 + \imath \langle r_c \rangle q -\frac{1}{2} \langle
r_c^2 \rangle q^2...$

   where $\langle r_c^R \rangle = \int^\infty_{-\infty} dr_c P_c r_c^R$ is a
constant for the Rth moment of $r_c$.

Then $ln [ \int^\infty_{-\infty} (P_c dr_c)e^{-\imath qr_c}dr_c]^R
= R ln [1 + \imath \langle r_c \rangle q - \frac{1}{2} \langle
r_c^2 \rangle q^2...] $

Using Taylor's series for $y\ll 1$: $ln (1+y) = y - \frac{1}{2}
y^2...$

$ln [ \int^\infty_{-\infty} (P_c dr_c)e^{-\imath qr_c}dr_c]^R = R[
\imath \langle r_c \rangle q - \frac{1}{2} \langle r_c^2 \rangle
q^2 - \frac{1}{2} (\imath \langle r_c \rangle  q)^2...]$

$ = R  [ \imath \langle r_c \rangle - \frac{1}{2} (\langle r_c^2
\rangle - \langle r_c \rangle ^2)q^2...]$

$= R [\imath \langle r_c \rangle - \frac{1}{2} (\Delta r_c)^2
q^2...]$ where $(\Delta r_c)^2 = (\langle r_c^2 \rangle - \langle
r_c\rangle ^2)$

Thus $ln [ \int^\infty_{-\infty} (P_c dr_c)e^{-\imath qr_c}dr_c]^R
= e^{\imath R \langle r_c \rangle q - \frac{1}{2} R
 \langle (\Delta r_c)^2 \rangle   q^2}$

And  $P_L(rst') =\frac{1}{2\pi} \int^\infty_{-\infty} e^{\imath R
\langle r_c \rangle q - \frac{1}{2} R
 \langle (\Delta r_c)^2 \rangle   q^2} dq$

From Integral tables  $P_L(r) = \frac{1}{\sqrt{2{\pi}}\sigma_r}
  e^-\frac{(r-r_0)^2}{2\sigma_r^2}$

where $r_0 = R \langle r_c \rangle $ and $\sigma_r = R \langle
\Delta r_c \rangle ^2$ and $\frac{r}{|T|}=\frac{1}{(st')} = k
\frac{dx}{t^* 2\pi}$

Then $e^-\frac{(r-r_0)^2}{2\sigma_r^2} = e^-\frac{(|T|k
\frac{dx}{t^* 2\pi}-R\langle r_c \rangle)^2}{2R (\langle
(r_c)^2\rangle - \langle r_c\rangle^2)} =
e^-\frac{(k-k_0)^2}{2\sigma_k^2}$

This allows us to define a function in k which we can label the
`probability function' that associates a probability with a
complex vector:

We then sum all values of $P_L(rst')$ for all values of t,
associated
with values of the vector position in time which we can define as a `probability function'...\\

 \beq \psi (x) =\int^\infty_{-\infty}\frac{1}{\sqrt{2{\pi}}\sigma_k}e^{-\frac{1}{2}
\frac{(k-k_0)^2}{(\sigma_k)^2}} e^{-\imath kx}dk \eeq

We now consider the probability of NOT having an interaction at
any of the previous times $(r-1)st',(r-2)st'$,...

For an IFE disturbance starting from an initial time magnitude
$|T| =0$, to calculate the probability of an interaction at
spatial position x each of the temporally precedent spatial
positions where an interaction did NOT occur: NOT$(n-1)dx$,
NOT$(n-2)dx$... must be considered, where there could have been
but was no interaction.

For a probability of interaction in space that was identically and
uniformly distributed in a one dimensional line this would be
straightforward: the n possible positions could be examined - each
separated by a very small distance $dx$ prior to the interaction
at x.

The probability of interaction in a very short space $dx$ can be
defined as $(B dx)$ where B is the probability density (of an
interaction with another IFE disturbance).

So the probability of non-occurrence in a very short space is
$(1-Bdx)$

If a distance x $=n dx$ is travelled before an interaction then
where $P_N(x)$ is the probability  for no interaction up to x:
$P_N(x) = (1-B dx)^{n} $

i.e. $P_N (x) = (1-B dx)^{n} = (1-\frac{Bx}{n})^n$

For a large x then $n=\frac{x}{dx}\longrightarrow\infty$. i.e. it
might at first be expected:

\beq  P_N (x) = e^{-Bx} \eeq

However, B, the probability density of an interaction in each
short spatial position varies according to the number of
alternative State positions at each possible Space position x. The
range of possible \textit{State positions itself} will vary at
different spatial positions.

Assume that for each occasion that the IFE disturbance moves from
one State position to another or from one spatial position to
another there is a uniform (arbitrary) probability A of
interaction with another (group of) IFE disturbance (that depends
on the state of the other IFE disturbance).

To calculate the probability for an interaction at a specific
State position (rh) at a spatial position x all of the
probabilities for each possible State
position at x can be summed (see Diagram 2 below).\\

\hspace* {-3mm}DIAGRAM 2: possible State positions for interaction
at spatial position x

\hspace* {-11mm}$r_{x}t'$
\\[-1.6mm]
 \vdots\hspace* {7mm}*
\\[-1.6mm]
 \vdots\hspace* {8mm}.
\\[-1.6mm]
\vdots\hspace* {8mm}.
\\[-1.6mm]
\vdots\hspace* {8mm}.
\\[-1.6mm]
\vdots\hspace* {8mm}.
\\[-1.6mm]
\vdots\hspace* {8mm}.
\\[-1.6mm]
\vdots\hspace* {8mm}.
\\[-3.2mm]
\ldots\ldots\ldots\ldots\ldots\ldots \hspace* {8mm}\\[-1.6mm]\hspace* {10mm}x\\

For an interaction to occur at State position $(r_xh)$ at a given
spatial position x, there must have been no interactions at each
of the previous possible \textit{and temporally precedent} State
points ($r_x-1)h$, $(r_x-2)h$...etc.

To calculate the probability of an interaction at a particular
State position a similar method can be used to that initially
assumed for spatial position.

Define probability of an interaction at a State position$ = A$

$P_N(rst') = (1- A dst')^r = (1-A \frac{rst'}{r})^r = e^{-A rst'}$

The probability of NOT having an interaction up to time $|rst|$
is:

\beq P_N(rst'/x=0) = e^{-Arst'} \eeq

We can calibrate between A and B: $ P_N(rst') = e^{-Brst'
\frac{dx}{t^*}}$

It is straightforward to calculate mean and variance using
this\footnote{For simplicity let $l = \frac{\imath rst'}{\imath
r_x st'}$. For a function B(x) mean $<x>=\int^\infty_{-\infty}x
B(x)dx$
   $$ \textnormal{i.e. } mean = l_o = <l>=\int^\infty_{-\infty}lAe^{-lA}dl  =\frac{A}{A^2} = \frac
   {1}A
   $$

$$   variance = \sigma^2_l=<l^2> - (<l>)^2$$

$$ <l^2> =\int^\infty_{-\infty}l^2(A)e^{-lA}dl = \frac{2A}{A^3} =
\frac {2}{A^2} $$

$$ i.e.~ \sigma^2_l=\frac {1}{A^2}$$}.

However, further alternative possible \textit{spatial} positions
such as at x = (n-1)d, (n-2)d...etc must be covered.

The total time magnitude $|T|$ can be composed in more than one
way (through variations in State and Space positions). Therefore
for a particular State position (rh) not only non-occurrences at
(r-1)h, (r-2)h..., but also for each of these State positions, the
non occurrences at all the coterminous \textit{spatial} positions
\textbf{\textit{which provide the same time magnitude $|T| =
|\imath rst'|$}} must be considered.

To establish $P(|T|)$, the probability of an interaction in a time
$|T|$, all the ways in which $|T|$ can be formed from the
combination of the first spatial position, the second spatial
position etc...must be calculated.

For more than one possible Space positions, all possible State
positions also must be accounted for at the second Space position
which in combination with the beta-time (caused by the movement in
spatial position) can comprise the same time magnitude equal to
$|\imath rst'|$ in the first Space position.

These possible combinations of certain State position (rh) and
specific Space position x (=nd) potentially exist only for those
combinations which have the same time magnitude $|T| = |\imath
rst'|$ such that $|T| = \sqrt{(nt^{*})^2 + (npst' + rst')^2 } $
where r is the State position that can occur at any Space
position.

For each possible interaction at a Space position x and State
position (rh), all possible non-interactions must be covered for
State and Space positions on an associated temporal arc. To
calculate possible positions are on this arc, a fundamental point
lattice calculation originated by Gauss (see Appendix) can be
used. This shows that $C(|T|)$, the number of exactly permissable
(integer) points on a temporal arc that can compose a time
magnitude $|T|$, is:

\beq C(|T|) = 2\pi |T| \eeq

For a particular State position (rh), not only all of the
potential interactions that did \textit{not} occur at alpha-times
$(r-1)\imath st'$, $(r-2)\imath st'$...must be accounted for.
Additionally all of the feasible interactions that could have, but
did not occur at alternative beta-times must be included - such as
$(|\imath rst'-t^*|),(|\imath rst'-t^* -1|)...$ at a second
\textit{spatial} position - and further $(|\imath
rst'-nt^*|),(|\imath rst'-nt^* -1|)...$ at the nth spatial
position.

\textbf{\textit{Calculation of the probability of NON-interactions
requires summation of the area of the arc of every possible State
position at every possible spatial position}}

The mechanics for this calculation are facilitated by working
backwards and investigating historically the non-occurrences of
interactions for Space and State positions.

A convolution method enables aggregation of all the possible
probabilities. To illustrate this, we can first calculate
notionally for two spatial positions only.

We sum for every State position against all the non-events at all
possible State positions.

From equation (6) and using (7) to locate the exactly permissable
(integer) values only:

$$ P_N(rst')  = \int^\infty_{-\infty}A e^{-Ar'} . A e ^{A(r-r')}dr'  $$

with the constraint that r' and (r-r') are not negative - i.e. $A
e^{-Ar'}$ and $A e ^{A(r-r')}$ can be represented as H(r') and
H(r-r') (Heaviside step functions):

$$ \int^\infty_{-\infty}A e^{-Ar'}H(r') . A e ^{A(r-r')}H(r-r')dr'  $$

$$ = \int^r_0 A e^{-Ar'} . A e ^{A(r-r')}dr'   = A^2 re^{-Ar}$$

Similarly using this last result for 3 positions:

$$ \int^\infty_{-\infty}A^2 re^{-Ar}H(r') . A e ^{A(r-r')}H(r-r')dr'  $$

$$ = \int^r_0 A^2r e^{-Ar'} . A e ^{A(r-r')}dr'  = \frac{A^3}{2} r^2e^{-Ar}$$

And for all possible n positions across the temporal arc, through
inference:

\beq P(r) = \frac{A^n r^{n-1}}{(n-1)!}e^{-Ar} \eeq

Note that (as with the earlier calculations for Mean and
Variance):
 \beq Mean = r_o = <r>=\int^\infty_{-\infty}\frac{(rA)^n}{(n-1)!}e^{-rA}dr  =\frac{n}{A}  \eeq

\beq   \textnormal{similarly }Variance = \sigma^2_r = \frac
{n}{A^2} \eeq

let $a = \frac{Ar}{n}$ , then from (8)

 $$P(r) = \frac {A n^{n-1}} {(n-1)!}
a^{n-1}e^{-na} = \frac{An^n}{n!}a^{n-1}e^{-na}$$

If we replace with z = a-1 then

$$ P(r) = \frac {An^n} {n!} (1+z)^{n-1}e^{-n(1+z)}$$

Assuming that n is large, P(r) can be expressed more conveniently
using Stirling's factorial expansion:\footnote{See Jeffreys (1)}

 $$ n! = \sqrt{(2{\pi}n)}n^n e^{-n}$$

Then $$P(r) = \frac {A}{\sqrt{2{\pi}n}}(1+z)^{n-1}e^{-nz}$$

But $e^{-nz} = 1 -\frac{nz}{1!}+\frac{(nz)^2}{2!}...$

And from binomial
expansion:~$(1+z)^{n-1}=1+(n-1)z+\frac{(n-1)(n-2)z^2}{2}+...$

Then collecting polynomials: $$ P(r) =
\frac{A}{\sqrt{2{\pi}n}}(1-z-\frac{1}{2}(n-2)z^2+...)$$

Ignoring $\frac{1}{n}$ for large n denominators and using the
above series for $e^{-nz}$:

$$ P(r)= \frac{A}{\sqrt{2{\pi}n}}e^{\frac{1}{2}n(z-\frac{1}{n})^2} =\frac{A}{\sqrt{2{\pi}n}}e^{-\frac{1}{2}n(a-1-\frac{1}{n})^2} $$

Substituting back for $a = \frac{Ar}{n}$

$$ P(r) =\frac{A}{\sqrt{2{\pi}n}}e^{-\frac{1}{2}n (\frac{Ar-1}{n}-1)^2}   =\frac{A}{\sqrt{2{\pi}n}}e^{-\frac{1}{2}\frac{(Ar-(n+1))^2}{n}}$$

For large n, $(n+1)\sim n$ and:

$$ P(r) =\frac{A}{\sqrt{2{\pi}n}}e^{-\frac{1}{2}\frac{(Ar-n)^2}{n}} =\frac{A}{\sqrt{2{\pi}n}}e^{-\frac{1}{2}
\frac{(r-\frac{n}{A})^2}{\frac{n}{A^2}}}$$

But from (9) and (10) $r_0 = \frac{n}{A}$ and $\sigma_r^2 =
\frac{n}{A^2}$

\beq P_N(r) =\frac{1}{\sqrt{2{\pi}}\sigma_r}e^{-\frac{1}{2}
\frac{(r-r_0)^2}{(\sigma_r)^2}}\eeq

This expresses the probability of a specific interaction at a
specific state position but does not account for the spatial
location.

As before  $e^-\frac{(r-r_0)^2}{2\sigma_r^2} =
e^-\frac{(k-k_0)^2}{2\sigma_k^2}$

Which allows us to define a further probability function in k:

 \beq \psi (x) =\int^\infty_{-\infty}\frac{1}{\sqrt{2{\pi}}\sigma_k}e^{-\frac{1}{2}
\frac{(k-k_0)^2}{(\sigma_k)^2}} e^{-\imath kx}dk \eeq

Thus from earlier $P(x)= P_L(x) P_N(x)$ and

\beq P(x)= \psi(x) \psi^*(x) \eeq

The interplay between P(k) and $\psi(x) = FT(P(k))$ produces a
property of the differential of P(x) indicated by P'(x):

$$\int^\infty_{-\infty}P'(x)e^{-ikx}dx = e^{-ikx}P(x)|^\infty_{-\infty} +ik\int^\infty_{-\infty}P(x)e^{-ikx}dx$$

And because $P(x)=e^{-x^2}$ and $e^{-\imath kx}$ is an oscillating
function:

\beq \int^\infty_{-\infty}P'(x)e^{-ikx}dx = ikFT(P(k))\eeq

Through a combination of such probability functions - say P(k) and
another similar probability function in k Q(k) - we can establish
an interesting relationship between the square of their product
and the product of their squares (see Rae 2002) since the integral
of a magnitude must always be positive:

$$\int^\infty_{-\infty}\|P(k)\{ \int^\infty_{-\infty}|Q(k)|^2dk \} - Q(k)\{ \int^\infty_{-\infty}P(k)Q^*(k)dk \}\|^2 ]dk\geq 0$$

Expanding this squared magnitude as the product of a function and
its conjugate:

$\int^\infty_{-\infty}[P(k)\{ \int^\infty_{-\infty}|Q(k)|^2dk \} -
Q(k)\{ \int^\infty_{-\infty}P(k)Q^*(k)dk \}]
\newline\hspace*{20mm} [P^*(k)\{ \int^\infty_{-\infty}|Q(k)|^2dk
\} - Q^*(k)\{ \int^\infty_{-\infty}P^*(k)Q(k)dk \}] dk \geq 0$
\\
\\
Multiplying out the square brackets obtains:

\beq \{\int^\infty_{-\infty}P(k)Q(k)dk \}^2 \leq
\int^\infty_{-\infty}|P(k)|^2dk \int^\infty_{-\infty}|Q(k)|^2dk
\eeq

For the variance of x and k $\sigma_x^2 = \langle x^2 \rangle
-\langle x \rangle ^2 $ and $\sigma_k^2 = \langle k^2 \rangle -
\langle k \rangle ^2$ and assuming that $\langle x\rangle =\langle
k\rangle =0$ \footnote {For the case where $\langle x\rangle \neq
0$ then we can perform a displacement function such that $\langle
x'\rangle =0$ and it can be shown (e.g. Jeffreys 1939) that the
product $\sigma_x^2\sigma_k^2$ then remains the same as for
$\langle x\rangle =\langle k\rangle=0$.} we form the product:

$$ \sigma_x^2\sigma_k^2 = \int^\infty_{-\infty}x^2|P(x)|^2dx
\int^\infty_{-\infty}k^2 |(P(k))|^2dk $$

However, we can show that:\footnote {This is the ``Parseval''
identity:

$$\int^\infty_{-\infty}P(k) P^* (k) dk = \int^\infty_{-\infty}
P(k)\{\int^\infty_{-\infty}FT(P^*(k))e^{\imath rk}\}dk$$

$$ =\int^\infty_{-\infty}\int^\infty_{-\infty}P(k)e^{\imath rk} P^*
(k)dk$$

$$ =\int^\infty_{-\infty} FT(P(k))
FT^*(P(k))$$ }

$$\int^\infty_{-\infty}|(P(k))|^2dk =
\int^\infty_{-\infty}|FT(P(k))|^2dk$$

$$ \textnormal{Hence } \sigma_x^2\sigma_k^2=
\int^\infty_{-\infty}|xP(x)|^2dx.\int^\infty_{-\infty}|ikFT(P(k)|^2dk$$

From (15):

$$ =
\int^\infty_{-\infty}|xP(x)|^2dx\int^\infty_{-\infty}|P'(x)e^{-\imath
xk}|^2dx =
\int^\infty_{-\infty}|xP(x)|^2dx\int^\infty_{-\infty}|P'(x)|^2dx$$

From (16) we have $\{\int^\infty_{-\infty}P(k)Q(k)dk \}^2 \leq
\int^\infty_{-\infty}|P(k)|^2dk \int^\infty_{-\infty}|Q(k)|^2dk$

$$ \sigma_x^2\sigma_k^2\geq\int^\infty_{-\infty}|x(P(x) P^{*'}(x)dx|^2 $$

$$ \sigma_x^2\sigma_k^2\geq\int^\infty_{-\infty}|\frac{x}{2}\frac{d}{dx}|(P(x))|^2dx|^2 $$

$$ \sigma_x^2\sigma_k^2\geq\frac{1}{4}\int^\infty_{-\infty}{|P(x)^{2}dx|}^{2} $$

And since $\int^\infty_{-\infty}|P(x)^2dx|^2$ is the probability
of finding the IFE disturbance \textit{anywhere} $ = 1$. Then:

\beq \sigma_x\sigma_k \geq\frac{1}{2} \eeq

\section{Mass and Momentum}

The velocity $v = \frac{d}{\sqrt{(t^* )^2 + (pst')^2 }}$ of an IFE
disturbance moving away from a notional fixed reference point can
be combined with that of another IFE disturbance moving away in
the opposite direction from the fixed reference point at a
velocity $u = \frac{d}{\sqrt{(t^* )^2 + (qst')^2 }}$. This
produces a calculation for the resultant velocity from two
independent velocities.

In time $\sqrt{(t^* )^2 + (pst')^2}$ the distance D travelled by
both disturbances is:

$ D = d + \frac{d}{\sqrt{(t^* )^2 + (qst')^2 }}\sqrt{(t^* )^2 +
(pst')^2}$

However, during the period of time $\sqrt{(t^* )^2 + (pst')^2}$
which accounts for a movement in space dx for the first IFE
disturbance, the number of \textit{beta-time} increments must be
accounted for by the second IFE disturbance (determined by its
State$\rightarrow$Space trigger-point qst') which may overlap with
those of the first.

To establish how many ``extra" incidents of beta-time$(t^*)$ occur
in this time, in a theoretical amount of time stretching across
$\sqrt{(t^* )^2 + (pst')^2}\sqrt{(t^* )^2 + (qst')^2 }$ there will
be an extra number N of incidents of $t^*$ where:

$ N = \sqrt{(\sqrt{(t^* )^2 + (pst')^2} + \sqrt{(t^* )^2 +
(qst')^2})^2 - \{(\sqrt{(t^* )^2 + (pst')^2})^2 + (qst')^2}\} $

This gives a \textit{rate} of discrepancy of extra $t^*$ per unit
of time such that:

$rate = \frac{\sqrt{(\sqrt{(t^* )^2 + (pst')^2} + \sqrt{(t^* )^2 +
(qst')^2})^2 - \{(\sqrt{(t^* )^2 + (pst')^2})^2 + (qst')^2} \}
}{\sqrt{(t^* )^2 + (pst')^2}\sqrt{(t^* )^2 + (qst')^2 }}$

In an amount of time $\sqrt{(t^* )^2 + (pst')^2}$ there will be
$\frac{\sqrt{(t^* )^2 + (pst')^2}}{\sqrt{(t^* )^2 + (qst')^2 }}$
opportunities for an extra ``skip" of beta-time.

The total number of extra incidents of $t^*$ will be:

 $\frac{\sqrt{(t^* )^2 + (pst')^2}}{\sqrt{(t^* )^2 + (qst')^2
}}\frac{\sqrt{(\sqrt{(t^* )^2 + (pst')^2} + \sqrt{(t^* )^2 +
(qst')^2})^2 - \{(\sqrt{(t^* )^2 + (pst')^2})^2 +
(qst')^2}\}}{\sqrt{(t^* )^2 + (pst')^2}\sqrt{(t^* )^2 + (qst')^2
}}$

Then the amount of time t we have to consider when calculating the
combined velocity of the two IFE disturbances is:

\hspace*{-15mm} $t = \sqrt{(\sqrt{(t^*)^2 + (pst')^2})^2 + (t^*)^2
[\frac{\sqrt{(t^* )^2 + (pst')^2}}{\sqrt{(t^* )^2 + (qst')^2
}}\frac{\sqrt{(\sqrt{(t^* )^2 + (pst')^2} + \sqrt{(t^* )^2 +
(qst')^2})^2 - \{(\sqrt{(t^* )^2 + (pst')^2})^2 +
(qst')^2}\}}{\sqrt{(t^* )^2 + (pst')^2}\sqrt{(t^* )^2 + (qst')^2
}}]}$

$ = \sqrt{(\sqrt{(t^*)^2 + (pst')^2})^2 + (t^*)^2
[\frac{2(\sqrt{(t^* )^2 + (pst')^2}}{\sqrt{(t^* )^2 + (qst')^2 }}
+ \frac{(t^*)^2}{(t^*)^2 + (qst')^2}} ]$

$ = \sqrt{(t^*)^2 + (pst')^2} + \frac{(t^*)^2}{\sqrt{(t^* )^2 +
(qst')^2 }} + 2(t^*)^2 \frac{\sqrt{(t^* )^2 +
(pst')^2}}{\sqrt{(t^* )^2 + (qst')^2}} $

$ = [ \sqrt{\sqrt{(t^*)^2 + (pst')^2} + \frac{(t^*)^2}{\sqrt{(t^*
)^2 + (qst')^2}}} ]^2$

$= \sqrt{(t^* )^2 + (pst')^2} + \frac{(t^*)^2}{\sqrt{(t^* )^2 +
(qst')^2}} $

Then the combined velocity V of the two IFE disturbances is:

\beq  V = \frac{ d + \frac{d}{\sqrt{(t^* )^2 + (qst')^2
}}\sqrt{(t^* )^2 + (pst')^2}}{\sqrt{(t^* )^2 + (pst')^2} +
\frac{(t^*)^2}{\sqrt{(t^* )^2 + (qst')^2}}} \eeq

Consider two IFE disturbances of equal rest mass $m_{0}$ and equal
velocity $u$ colliding in a non-elastic way from opposite
directions (say a mass moving from the left and a mass moving from
the right), resulting in a stationary object of mass $M_{0}$.
Suppose that mass is not necessarily constant so that the moving
mass $m_u$ may be different from the stationary rest mass $m_{0}$.

From the perspective of the second IFE disturbance mass moving
from the right then the disturbance moving from the left has an
effective velocity V (of the combined velocities) and a mass
$m_V$. If it then hits the second disturbance of mass $m_{0}$ this
results in an IFE disturbance of mass $M_u$ moving with a velocity
u.

From equation $18$ the effective velocity of two combined equal
velocities each moving towards one another with velocity $u=
\frac{d}{\sqrt{(t^* )^2 + (pst')^2}}$ is:

\beq V = \frac{2d\sqrt{(t^* )^2 + (pst')^2}}{2(t^* )^2 + (pst'
)^2} \eeq

Employing two fundamental laws (through empirical experience):

 (1) Conservation of Momentum  i.e. $m_V V = M_u u$

 (2) Conservation of Mass i.e. $m_V + m_{0} = M_u$

Combining these two conservation laws and eliminating $M_u$:

 \beq \frac{m_V}{m_{0}}= \frac{u}{V-u} \eeq

 Making use of $u= \frac{d}{\sqrt{(t^* )^2 + (pst')^2}}$ and equation 18:

$$ \frac{m_V}{m_{0}} =  \frac{\frac{d}{\sqrt{(t^* )^2 +
(qst')^2}}}{\frac{2d\sqrt{(t^* )^2 + (qst')^2}}{2(t^* )^2 + (pst'
)^2} -\frac{d}{\sqrt{(t^* )^2 + (qst')^2}}}=
\frac{d}{\frac{2d\{(t^* )^2 + (qst')^2\}}{2(t^* )^2 + (pst' )^2}
-d}$$

\beq \textnormal{Then }\frac{m_V}{m_{0}}= \frac{2(t^*)^2 +
(pst')^2}{(pst')^2} \eeq

If we multiply by $ \frac{d^2}{(t^*)^2}$:

\beq m_V\frac{d^2}{(t^*)^2} = m_{0}\frac{d^2}{(t^*)^2} +
2m_{0}\frac{d^2}{(pst')^2} \eeq

The second expression on the right indicates a multiple of the
rest mass with some form of the square of the velocity.

The traditional Newtonian formulation of kinetic energy is
$\frac{1}{2}m_{0}V^2$ and from eq (19) $V = \frac{2d\sqrt{(t^* )^2
+ (pst')^2}}{2(t^* )^2 + (pst' )^2}$. For speeds much less than
the speed of light $\frac{1}{2}m_{0}V^2 \sim \frac{2m_0
d^2}{(pst')^2}$. This is the last expression on the right of
equation (22) which suggests that equation $22$ refers to the
\textit{energy} of the IFE disturbance, where the term
$2m_{0}\frac{d^2}{(pst')^2}$ indicates its kinetic energy.
Consequently for the rest energy $E_{0}$ of the disturbance:

\beq E_{0} = m_{0}\frac{d^2}{(t^*)^2}  \eeq

And for the total energy $E_{T}$ of the IFE disturbance:

\beq E_{T} = m_V \frac{d^2}{(t^*)^2} \eeq

These are, of course, instances of Einstein's familiar expression
$E = mc^2$.

Since $E_{T} = \frac{h}{st'}$ then \footnote{see earlier paper
Brown (2003)} from (22):

\beq  m_0  =\frac{h(t^*)^2}{d^2(st')}\{\frac{(pst')^2}{(pst')^2 +
2(t^*)^2 }\} \eeq

A term can be defined in line with kinetic energy and designated
``pure kinetic mass" $m_k$:

\beq m_k=m_V - m_0 =
\frac{h(t^*)^2}{d^2(st')}\{\frac{2(t*)^2}{(pst')^2 + 2(t^*)^2 }\}
\eeq

A neat balance is therefore exhibited. When (pst') is very large
compared to $t^*$ then rest mass will dominate the total mass
(i.e. $m_0 \rightarrow\frac{h(t^*)^2}{d(st')}$) and the pure
kinetic mass will be negligible. As (pst') decreases, however,
then the proportion of pure kinetic mass to rest mass will
increase. Ultimately, for a photon - which has no trigger point -
then (pst') is $0$ and its mass comprises pure kinetic mass only:
it has no rest mass.

An interesting formulation for the total energy can be obtained
noting that:

$ (m_V V)^2 c^2= \frac{4h^2(t^*)^2\{(t^*)^2 +
(pst')^2\}}{(st')^2\{2(t^*)^2 + (pst')^2\}}$ and $(m_0)^2 c^4 =
\frac{h^2(t^*)^4 (pst')^4}{(st')^2\{2(t^*)^2 + (pst')^2\}}$. Thus

\beq (m_V V)^2 c^2 + (m_0)^2 c^4 = \frac{h^2}{(st')^2} = (e_T)^2
\eeq

This implies that the total energy squared is equal to sum of the
momentum squared multiplied by the speed of light squared and the
the rest energy squared. This occurs once again because we have to
consider what a moving ``particle" is: for which there will only
be a series of state changes which are included in the calculation
of the time. Whereas from the perspective of another IFE
disturbance this disturbance is moving and the time taken for
movement in Space must also be considered.

Viewing energy as rate of change of State, the \textit{perception}
of the magnitude of this quantity will vary from different
vantages moving at different velocities. From the perspective of a
mass moving at speed V its pure kinetic mass can be isolated from
a vantage point moving at the same speed as the pure kinetic mass
(i.e. we can neglect the inertial rest mass). The pure kinetic
mass itself comprises a moving IFE disturbance moving from a
notional fixed point with a certain speed. This moving IFE
disturbance could assume a range of speeds with respect to the
possible speeds of the fixed point, whilst nevertheless
maintaining the effective speed V. Average quantities enable
calculation of the speed of the fixed point and the speed of the
internal IFE disturbance which springs from it as both having the
same speed u.

If we now calculate the momentum, from (19) and (24):

$$ m_V V = \frac{2h}{(st')}\frac{(t^*)^2}{\{2(t^*)^2 + (pst')^2\}}\frac{\sqrt{(t^*)^2 + (pst')^2}}{d}$$

Yet from the above discussion this represents the product of the
pure kinetic energy (which from the perspective of an entity
moving at speed $u = \frac{d}{\sqrt{(t^* )^2 + (pst')^2}}$ is its
total energy) of the moving mass and the inverse $\frac{1}{u}$ of
the internal IFE disturbance velocity. Hence, expressing total
energy using s' from the point of view of the moving entity $E_T =
\frac{h}{s't'}$, and using $u=\frac{\lambda}{(s't')}$:

\beq \textnormal{Momentum }P = m_V V = \frac{h}{\lambda} \eeq

Since $ P  = \frac{h}{\lambda} = \frac{hk}{2\pi}$ then from (17)

\beq \sigma_x\sigma_p \geq\frac{h}{4\pi}\eeq

This is the familiar expression of Heisenberg's Uncertainty
Principle.

\section{Conclusions}

A mechanism for formalising the statistical underpinnings of
quantum calculations provides both a means for calculation and a
rationale for the quantum uncertainty of position and momentum. A
later paper is intended on the application of this method to the
theory of gravity. Detailed computer models and discussion are
available from the author on request.

\section{Appendix: The number of potential positions precisely lying
on the temporal arc}

The goal is to locate the number of \textit{potential} positions
on the temporal arc formed through the time magnitude $|T| =
\sqrt{(nt^{*})^2 + (npst' + rst')^2 }$. Since t* and t' are finite
numbers, and since n, p, s and r are integers then only a small
subset of positions on the temporal arc can exist to form $|T|$.
Since this can effectively be represented as the root of a sum of
two squares, then we effectively want to estimate the number of
lattice points $C(|T|)$ on the circumference of a circle of radius
$|T|$.

A theory of point lattices can determine the number of possible
lattice points \textit{in and on} a circle C($|T|$) of radius
$|T|$. If we consider the circle at the origin of a fundamental
point lattice with each lattice point as the centre of a unit
square with sides parallel to the axes t* and t', then the area of
all the squares whose \textit{centres} are inside or on $C(|T|)$
can be analysed. This area $L(|T|)$ comprises a number of complete
squares entirely within the circle, and also \textit{a number of
squares that are divided by the circle of radius $|T|$}

Some parts of squares with \textit{centres} inside the circle of
radius $|T|$ will remain outside of the circumference. Equally
some squares with \textit{centres} outside the circle have
boundaries fitting partly within the circle's perimeter. If we
theoretically shade in all the complete squares whose centres are
in or on the circle, then we can bound the shaded area $L(|T|)$
from below and above - we find the largest disk whose interior is
completely shaded, and the smallest disk whose exterior is
completely unshaded. Since the diagonal of a unit square is
$\sqrt{2}$ then all shaded squares must be contained in a circle
of radius $ = |T| + (\sqrt{2}/2)$. Similarly the circle whose
radius $ = |T| - (\sqrt{2}/2)$ is contained entirely within the
shaded squares. Consequently

$ \pi(|T|^2 -\sqrt{2}|T| - \frac{1}{2}) \leq \pi(|T|^2
-\sqrt{2}|T| + \frac{1}{2}) \leq L(|T|) \leq \pi(|T|^2 +
\sqrt{2}|T| + \frac{1}{2}) $

Which implies that

$ |\frac{L(|T|)}{|T|^2} - \pi | \leq \pi (\sqrt{\frac{2}{|T|^2}}
+\frac{1}{2|T|^2}) $

Since $(\sqrt{\frac{2}{|T|^2}} +\frac{1}{2|T|^2})$ tends to 0 as $
|T|\rightarrow\infty$ then $ L(|T|)/|T|^2 \rightarrow\pi $

i.e. $L(|T|) = \pi |T|^2$.

This defines the number of lattice points both in and on a circle
of radius $|T|$. The number of points solely \textit{on} the
circle of radius $|T|$ is simply $C(|T|) = 2\pi|T|$.

Whilst different arcs will have volatile numbers of potential
compositions through nt* and rst'(and some arcs will be
effectively prime composed through only a single instance of n and
rst') an average value for the number of possible positions on a
variable temporal arc, will be effective if summed over a
large/infinite series - probabilities will be summed. We therefore
sum the first n values of $L(|T|)$ (the number of possible lattice
positions on a circle of radius $|T|$) and divide by n to obtain
an associated average for the total number of of ways for
combining the two axes of time to form the single time magnitude:

$ \frac{C(|T|)}{|T|} = \frac{C(0)+ C(1) + C(2) +...+C(|T|)}{|T|} $

Therefore $C(|T|) = 2\pi|T|$ can be utilised.

\section{Acknowledgements}
The assistance of University College London Computer Science
department is gratefully acknowledged.

\section{References}

\hspace* {5mm} (1) Brown D (2003): ``A Graph method for mapping
changes in temporal and spatial phenomena with relativistic
consequences" gr-qc/0312004

 (2) Jeffreys H (1939): ``Theory of Probability" ,
OUP.

(3) Gauss C F (1863-1933): ``Werke" , Gottingen: Gesellschaft der
Wissenschafften,

(4) Olds C D, Lax A, Davidoff G (2000): ``The Geometry of Numbers"
, The Mathematical Association of America

(5) Rae A I M (2002): "Quantum Mechanics", IOP Publishing, Bristol

\end{document}